\documentclass[a4paper, 10pt]{article}
\usepackage{inputenc}
\usepackage[top=2.5cm, left=2.5cm, right=2cm, bottom=3cm]{geometry}
\usepackage{amssymb}
\usepackage{amsthm}
\usepackage{newlfont}

\begin{document}

\vspace*{2cm}

\begin{center}
{\Large\textbf{General Relativity in two dimensions:}}

\medskip

{\Large\textbf{a Hamilton-Jacobi constraint analysis}}

\vspace*{1cm}

\emph{M.C.Bertin$^1$, B.M.Pimentel$^1$} and \emph{P.J.Pompeia$^2$}

\vspace*{1cm}

$^1$ Instituto de F\'{i}sica Te\'{o}rica UNESP - S\~{a}o Paulo State University.

Caixa Postal 70532-2, 01156-970 S\~{a}o Paulo, SP, Brazil.

\medskip

$^2$ Divis\~{a}o de Confiabilidade Metrol\'{o}gica Aeroespacial

Instituto de Fomento e Coordena\c{c}\~{a}o Industrial.

Pra\c{c}a Marechal Eduardo Gomes, 50, Vila das Ac\'{a}cias, 12.228-901

S\~{a}o Jos\'{e} dos Campos, SP, Brazil.

\textit{e-mail: pimentel@ift.unesp.br}
\end{center}

\vspace*{0.2cm}

\begin{abstract}
We will analyze the constraint structure of the Einstein-Hilbert first-order 
action in two dimensions using the Hamilton-Jacobi approach. We will be able to find a set of involutive, as well as a set
of non-involutive constraints. Using generalized brackets we will show how
to assure integrability of the theory, to eliminate the set of
non-involutive constraints, and to build the field equations.
\end{abstract}


\section{Introduction}

For some time the Hamilton-Jacobi (HJ) formalism for constrained systems \cite{gu1} is being developed based on the work of Carath\'{e}odory on
variational principles and theory of first-order partial differential equations \cite{cara}. The Carath\'{e}odory's approach is characterized by the
so called \textquotedblleft complete figure\textquotedblright\ of the variational calculus, showing the link between the stationary action
principle, the HJ partial differential equation, and the set of first-order Hamiltonian ordinary differential equations. The geometric implications of this point of view are quite powerful, since it permits the analysis of a very wider range of problems, both in mathematics and physics, than the
problems which concern classical mechanics, to what the original method of Carath\'{e}odory was bounded. Restricting ourselves to constrained systems, it is well established today how the HJ formalism is capable to describe Lagrangians with higher-order derivatives \cite{Pim1,Pim2}, and
Berezian systems \cite{Pim3}. Moreover, several applications have been made \cite{gu2,gu3,cor,Pim4,Pim5}.

In the HJ formalism the constraints of a singular system form a set of partial differential equations that must fulfill integrability conditions to
guarantee integrability. These integrability conditions are equivalent to the consistency conditions present in the Hamiltonian approach \cite{dirac, dirac1}, and separate the constraints in involutive and non-involutive under the Poisson Brackets (PB) operation. In \cite{FOA}, the authors showed that for Lagrangians linear on the velocities it is possible to define Generalized Brackets (GB) with which non-involutive constraints become in involution. In a more recent work \cite{NIC}, the integrability conditions are analyzed to show that the GB are a general structure, not restricted to first-order actions, solving the problem of integrability due to the presence of non-involutive constraints.

At least for theories that present only non-involutive constraints the HJ approach is completely equivalent to Dirac one. On the other hand,
systems with involutive constraints are still object of investigation on the HJ scheme, since they represent identically integrable systems with more than one evolution parameters (in the mathematical sense, more than one independent variables). Constraints in involution are actually
first-class constraints in the canonical formalism, so we expect that understanding the role of these constraints within the HJ method would open
a channel to the study of gauge theories. The HJ formalism can be a rich source of information, since its geometric interpretation is clear and
immediate, allowing the study of symmetries on several fronts of investigation.

Therefore, in this work we will study the first-order form of the Einstein-Hilbert Action (EHA) in two dimensions, which is a theory that
presents involutive and non-involutive constraints. This theory is well explored in the literature within the scope of the Hamiltonian formalism
\cite{grav1,grav2,grav3}, which will allow us to compare results of both approaches. Besides that, the analysis of the EHA in two dimensions is an
interesting subject by itself: several models in lower dimensions are shown to be soluble quantum systems \cite{Martinec,jackiw,Henneaux}, providing
new perspectives on the non-perturbative quantization in higher dimensions. Moreover, the interest in lower dimension gravity is also present in
string theory \cite{Rebbi,Polyakov,Schwarz,Teitelboim}. The canonical analysis made in \cite{grav1} shows a set of second-class constraints, as well
as a set of first-class constraints that closes an algebra with constant structure coefficients. The theory shows an $SO\left( 2,1\right) $ gauge symmetry and allows quantization by the Faddeev-Popov approach,
which is discussed by McKeon in \cite{McKeon}.

We will organize the paper as follows: in section 2 we will revise the HJ formalism for singular systems under the scope of Carath\'{e}odory. We will show how the GB can be defined in the presence of non-involutive constraints, taking as focus the integrability
conditions and its geometrical interpretation \cite{NIC}. In section 3 we will apply the analysis to the Einstein-Hilber action taking the metric and affine connection as independent fields. A more natural approach would be using the metric density instead, what is done in the canonical analysis in the mentioned literature. It happens, however, that the Palatini's descriptions in metric and metric density are not equivalent in two dimensions, since the metric density has only two independent components, while the metric approach presents three independent degrees of freedom. Although the structure is more complex, the metric-affine description deals naturally with this issue. Moreover, the Palatini's approach is also inequivalent to the purely metric description since the field equations are not sufficient to fix the connection to the Christoffel symbols \cite{Deser}. Despite of these issues we will be able to reproduce the algebra of the involutive constraints, and the field equations of the system at the classical level as well.

\section{HJ formalism revisited}

Let us revise the HJ formalism by the geometric point of view of Carath\'{e}odory's approach \cite{cara}, considering a system described by an
action and a Lagrangian function of $n$ variables $q^{i}$, $n$ velocities $\dot{q}^{i}$ and possibly of a \textquotedblleft time\textquotedblright\
parameter which we call simply as $t$. If the system obeys a minimum action principle, a possible solution will be also a solution of the
Euler-Lagrange equations, written as a curve $C:q^{i}=q^{i}\left( t\right) $, and the action calculated on that solution is minimum comparing to any
other curve in the neighbourhood. As a well behaved curve, we are able to define the velocities $\dot{q}^{i}=\phi \left( q,t\right) $, which are the
components of the velocity field tangent to $C$, as well as the canonical momenta $p_{i}\equiv \partial L/\partial \dot{q}^{i}$, which are the
coordinates of the cotangent space defined over $C$.

By using the equivalent Lagrangian method, Carath\'{e}odory found that there are two necessary and sufficient conditions for a minimum action on the
curve $C$. The first one is the existence of a family of surfaces in the configuration space, defined by a generating function $S\left( q,t\right)
=\sigma \left( t\right) $, which is orthogonal to the conjugate 1-form momentum $p=p_{i}dq^{i}$, \emph{ie} we must have $p=dS$, or, in components,
$p_{i}=\partial _{i}S$.

The second condition is that the function $S$ becomes a solution of the equation
\begin{equation}
\frac{\partial S}{\partial t}+\dot{q}^{i}\partial _{i}S-L=0.  \label{HJ1}
\end{equation}
This equation emerges as a condition of a point transformation with generating function $S$ such that the new Lagrangian function $\bar{L}=L-dS/dt$ is
zero on the solution. On the other hand, the first condition comes from the imposition $\partial \bar{L}/\partial \dot{q}^{i}=0$ also over $C$.

In order to write properly the above equation we must know the function $\phi \left( q,t\right) $. The best we can do without knowing the explicit solution of the system is to find equations of the type $\dot{q}^{i}=\eta ^{i}\left(
q,p,t\right) $, and this can be done if we are able to invert the relations $p_{i}\equiv \partial L/\partial \dot{q}^{i}$. This is possible only if the Hessian matrix $W_{ij}\equiv \partial p_{i}/\partial \dot{q}^{j}$ is non-singular. In this case equation (\ref{HJ1}) can be written as a first-order partial differential equation, known as the Hamilton-Jacobi equation. This system is called regular.

If the system is singular, \emph{ie} $\det W=0$, it is not possible to find all the velocities in the form $\dot{q}^{i}=\eta
^{i}\left(q,p,t\right) $, and the theory will present a number $k=n-m$ of canonical constraints
\begin{equation}
H_{z}^{\prime }=p_{z}+H_{z}=0,\ \ \ \ \ \ \ \ z=1,\ldots ,k,  \label{pconstr}
\end{equation}
where $k$ is the number of zero modes of the Hessian matrix, $m$ is the rank, and $H_{z}\equiv -\partial
L/\partial \dot{q}^{z}$. It means that $m $ velocities can be found in the form $\dot{q}^{a}=\eta ^{a}\left(
q,p,t\right) $, where $\left\{ a\right\} =\left\{ 1,\ldots ,m\right\} $. 

Wherever the constraints are valid, we are able to define the canonical Hamiltonian function by $H_{0}\equiv \dot{q}^{i}\partial _{i}S-L$, that will
not depend on the velocities $\dot{q}^{z}$ as well. Therefore, the HJ equation (\ref{HJ1}) has the same structure of (\ref{pconstr}). Defining $%
p_{0}\equiv \partial _{t}S$ as the canonical momentum related to the $x^{0}\equiv t$ variable we can write these equations in a unified way:%
\begin{equation}
H_{\alpha }^{\prime }=p_{\alpha }+H_{\alpha }=0,\ \ \ \ \ \ \ \ \alpha
=0,1,\ldots ,k.  \label{HJ}
\end{equation}
This system is a set of first-order partial differential equations, called the Hamilton-Jacobi partial differential equations (HJPDE).

It is most fortunate that the first-order form of these equations allows one to relate the HJPDE to a set of first-order total differential equations,
the characteristics equations (CE) of the theory. Two of them are given by%
\[dq^{a}=\frac{\partial H_{\alpha }^{\prime }}{dp^{a}}dx^{\alpha },\ \ \ \ \
dp_{a}=-\frac{\partial H_{\alpha }^{\prime }}{\partial q^{a}}dx^{\alpha }.\]
With these equations, which have a very familiar canonical form, we are able to show that the variables $\xi ^{\bar{a}}=\left( q^{a},p_{a}\right) $
are coordinates of a phase space with a non-degenerate symplectic structure, represented by the Poisson brackets%
\[\left\{ F,G\right\} =\partial _{A}F\ \Omega ^{AB}\ \partial _{B}G,\ \ \ \ \ \ \ \ \ \ A,B=1,\ldots ,m,m+1,\ldots ,2m.\]
We have defined the symplectic matrix $\Omega ^{AB}=\delta _{b}^{a}\left(\delta _{1}^{\zeta }\delta _{\sigma }^{2}-\delta _{\sigma }^{1}\delta
_{2}^{\zeta }\right) $, where $\zeta $ and $\sigma $ are 1 for coordinates and 2 for momenta.

The evolution of a function of this phase space is, then, given by%
\[dF\left( q^{a},p_{a}\right) =\left\{ F,H_{\alpha }^{\prime }\right\}dx^{\alpha }.\]
If $F$ depends on the variables $x^{\alpha }$ we must add to this equation the partial derivatives with respect to these variables. It is more
convenient for calculation to actually use the brackets below:%
\begin{equation}
\left\{ F,G\right\} =\partial _{I}F\ \Omega ^{IJ}\ \partial _{J}G,\ \ \ \ \ \ \ \ \ \ I,J=0,1,\ldots ,n,n+1,\ldots ,2n,  \label{PB}
\end{equation}%
which is a degenerate structure and must be used with caution. We can write, therefore,%
\begin{equation}
dF\left( q^{a},p_{a},x^{\alpha }\right) =\left\{ F,H_{\alpha }^{\prime
}\right\} dx^{\alpha }.  \label{ev}
\end{equation}%
From now on we will suppose all Poisson brackets to be defined as (\ref{PB}).

The most important feature of this structure is that it shows the role of the constraints as generators of a $\left( k+1\right) $-parameter family of
curves on the phase space. This family is the solution of the CE, which we can write as%
\begin{equation}
d\xi ^{A}=\left\{ \xi ^{A},H_{\alpha }^{\prime }\right\} dx^{\alpha }.
\label{CE1}
\end{equation}%
The parameters of the family are the variables $x^{\alpha }$, \emph{ie} the variables whose velocities could not be expressed are actually parameters
in equal status as $t$.

Another CE is given for the function $S$,%
\begin{equation}
dS=p_{a}dq^{a}-H_{\alpha }dx^{\alpha }.  \label{CE2}
\end{equation}%
This last equation can be solved by a quadrature if the solutions for equations (\ref{CE1}) are known. The form of the quadrature assumes the form
of a canonical action with several independent variables which corroborate the previous interpretation of the CE. We can compare the action%
\[I=\int Ldt,\]
to the integral form of the equation (\ref{CE2}),%
\[S=\int \left[ p_{a}dq^{a}-H_{\alpha }dx^{\alpha }\right] ,\]
and we find that the action $I$ is a solution of the HJPDE if%
\[L=p_{a}\dot{q}^{a}-H_{\alpha }\dot{x}^{\alpha }=p_{a}\dot{q}^{a}-H_{0}-H_{z}\dot{x}^{z}.\]
This relation can be written as%
\[L=p_{i}\dot{q}^{i}-H_{0}-H_{z}^{\prime }\dot{x}^{z},\]
by adding and subtracting a term $p_{z}\dot{q}^{z}$. Since $\dot{x}^{z}$ are not determined in the theory we are able to link this result to Dirac
formulation by considering these velocities as Lagrange multipliers and the primary Hamiltonian $H_{P}\equiv H_{0}+H_{z}^{\prime }\dot{x}^{z}$ as
generator of the time evolution. The CE (\ref{CE1}) can also be put in the form written by Dirac, $\dot{\xi}^{A}=\left\{ \xi ^{A},H_{P}\right\} $. At
this stage both formalisms are shown to be equivalent, although the HJ formalism presents a stronger theoretical link to the principle of least
action.

\subsection{The Complete Figure and Integrability Conditions}

Concerning the theory of partial differential equations, any set of first-order partial equations can be written with the help of a set of
vector fields that belongs to an affine space tangent to the family of surfaces which is the solution of these same equations. In the case of the HJPDE
we can define $k+1$ vector fields $X_{\alpha }\equiv \chi _{\alpha}^{I}\partial _{I}$, such that%
\begin{equation}
H_{\alpha }^{\prime }\equiv X_{\alpha }\cdot dS=\chi _{\alpha }^{i}\partial
_{i}S=\chi _{\alpha }^{i}p_{i}=0,  \label{xds=0}
\end{equation}%
in which $\chi _{\alpha }^{\beta }=\delta _{\alpha }^{\beta }$, and $\chi_{\alpha }^{a}=\left\{ q^{a},H_{\alpha }^{\prime }\right\} $ (only in this
equation we are considering the notation $i=\left( \alpha ,a\right) $). In this way not only the space of vectors $X_{\alpha }$ are tangent to the family, but
orthogonal to the momentum as well, as it would be expected. The tangent space of the vectors is actually isomorphic to the parameter space of the
variables $x^{\alpha }$, so we will make no distinction among them.

On the other hand, the CE can be written by%
\begin{equation}
d\xi ^{A}=\chi _{\alpha }^{A}dx^{\alpha },  \label{dx}
\end{equation}%
where $\chi _{\alpha }^{A}=\left\{ \xi ^{A},H_{\alpha }^{\prime }\right\} .$

It is immediate to see, by construction, that if the vector fields $X_{\alpha }$ form a complete basis on the parameter space, \emph{ie} if
they form a set of maximal linearly independent vector fields, the equations $X_{\alpha }\left( S\right) =0$ will be a complete set of partial
differential equations and the integrability is secured. Integrability is also necessary for the existence of a complete solution of the HJPDE. Linear
independence is fulfilled if, and only if%
\begin{equation}
\left[ X_{\alpha },X_{\beta }\right] \equiv X_{\alpha }X_{\beta }-X_{\beta
}X_{\alpha }=C_{\alpha \beta \gamma }X_{\gamma }.  \label{fic}
\end{equation}%
These are known as Frobenius' integrability conditions.

Let us remember that these vector fields are linear differential operators, and by applying the Frobenius' condition on $S$ we find that
$\left[X_{\alpha},X_{\beta }\right] \left( S\right) =0$ is also satisfied, since $X_{\alpha}\left( S\right) =H_{\alpha }^{\prime }=0$. The components
of $X_{\alpha }$ are related to the PB of the constraints so that the integrability conditions can be written in the form%
\begin{equation}
\left\{ H_{\alpha }^{\prime },H_{\beta }^{\prime }\right\} =C_{\alpha \beta
\gamma }H_{\gamma }^{\prime }=0.  \label{fic1}
\end{equation}%
Therefore, the constraints must form a system in involution with respect to the PB operation.

If a set of HJPDE is not integrable the reason may be that they are not complete, not linearly independent, or even both. In this case it is more
convenient to use the equivalent relations%
\begin{equation}
dH_{\alpha }^{\prime }=0  \label{int}
\end{equation}%
as the integrability conditions of the theory, as these conditions permits to detect linear dependence of the vector fields and to complete the set of HJPDE \cite{NIC}. On a general Lagrangian system the conditions (\ref{int}) can be of three types. The first type is the case in which some
constraints are in involution with the PB, and in this case the above integrability conditions are identically satisfied. But some of the constraints may result in equations of the type $f(q,x,p)=0$, which is the second type of condition. We are interested in these expressions at first place, because they must be considered as constraints in equality to the former set. They must also obey integrability conditions and may lead to other constraints. All possible constraints
should be found and inserted in the formalism as generators of the dynamics of the system.

Unlike the constraints (\ref{HJ}), the new constraints are not, in general, first-order equations, and there is not generally possible to relate
independent variables of the theory to these equations. To deal with this problem we may expand the parameter space with new arbitrary independent
variables, in such way that each new constraint generates a 1-parameter curve whose evolution parameter becomes the correspondent independent
variable. Let us suppose now that all constraints can be put in the form $H_{\alpha }^{\prime }=0$ with $\alpha $ covering all the expanded parameter
space. This set is supposed to be complete, and the new characteristics equations can be derived from the fundamental differential%
\begin{equation}
dF=\left\{ F,H_{\alpha }^{\prime }\right\} dx^{\alpha },  \label{FD}
\end{equation}%
when, now, $x^{\alpha }$ is the set of all parameters of the theory, including the ones related to the new constraints. By the same argument used
previously it can be shown that this procedure is equivalent to the definition of the extended Hamiltonian in Dirac approach.

As the third type of relation, it may happen that some of the integrability conditions give total differential equations that relate the differentials of the parameters. These relations indicate that the actual set of constraints has linearly dependent vector fields, which affects the integrability of the system as well. It is shown in the reference \cite{NIC} that these conditions lead to a generalized symplectic structure, the generalized brackets. The introduction of the GB as the structure that determines the dynamics of the system solves the problem of non-integrability due to non-involutive constraints, since, by construction, the Frobenius' condition is identically satisfied if we substitute the PB by the GB of the theory.

\subsection{Generalized Brackets}

Let us define the matrix $M$,
\begin{equation}
M_{\alpha \beta }\equiv \left\{ H_{\alpha }^{\prime },H_{\beta }^{\prime}\right\}.
\label{m1}
\end{equation}%
The integrability conditions $dH_{\alpha }^{\prime}=0$ assume the form
\[M_{\alpha \beta }dx^{\beta }=0.\]
If the parameters are considered independent, the only possible solution is given by $M_{\alpha \beta }=0$. However, it is possible that the
constraints do not obey the condition $M_{\alpha \beta }=0$, and in this case we have to consider linear dependence on the parameters $x^{\alpha }$
to fulfill the integrability conditions of the system.

Let us suppose the case in which $M$ has rank $m\leq n$. We should separate the parameter space in two subspaces: the first one
being of the coordinates $y^{a}$, related to the invertible sub-matrix $m\times m$. The second, the $n-m$ coordinates $x^{\bar{\alpha}}$ related to
the non-invertible part of $M$. The first set of integrability conditions gives
\begin{equation}
dy^{a}=-(M^{\prime -1})^{ab}\{H_{b}^{\prime },H_{\bar{\beta}}^{\prime }\}dx^{%
\bar{\beta}}.  \label{20}
\end{equation}%
In this equation $\bar{\beta}=0,1,\ldots ,n-m$, and $a=1,\ldots ,m$. On the other hand we can write the differential (\ref{FD}) as%
\begin{equation}
dF=\{F,H_{a}^{\prime }\}dy^{a}+\{F,H_{\bar{\beta}}^{\prime }\}dx^{\bar{\beta}%
}.  \label{21}
\end{equation}%
Using (\ref{20}),%
\begin{equation}
dF=\left[ \{F,H_{\bar{\beta}}^{\prime }\}-\{F,H_{a}^{\prime }\}(M^{\prime
-1})^{ab}\{H_{b}^{\prime },H_{\bar{\beta}}^{\prime }\}\right] dx^{\bar{\beta}%
}.  \label{22}
\end{equation}

The above equation motivates us to introduce the Generalized Brackets (GB)%
\begin{equation}
\{F,G\}^{\ast }\equiv \{F,G\}-\{F,H_{a}^{\prime }\}(M^{\prime
-1})^{ab}\{H_{b}^{\prime },G\},  \label{23}
\end{equation}%
In Dirac approach the non-singular submatrix $M^{ab}$ constructed in a similar way means that the constraints associated to this matrix are
second-class. In the HJ approach it means that the system is not integrable, because every constraint has
non-zero PB with at least one of the others. However, if we use the GB as the bracket that gives us the
dynamic of the system, we see that all constraints will have zero GB with each others and the theory becomes
integrable. The constraints become involutive with the GB. We will impose, then, that the evolution must be given by
\begin{equation}
dF=\{F,H_{\bar{\beta}}^{\prime }\}^{\ast }dx^{\bar{\beta}}.  \label{24}
\end{equation}%

For the remaining constraints, $H_{\bar{\beta}}^{\prime }$, we have the integrability conditions%
\begin{equation}
\{H_{\bar{\beta}}^{\prime },H_{0}^{\prime }\}^{\ast }=0.  \label{25}
\end{equation}%
So, we need to get a null GB between the constraints $H_{\bar{\beta}}^{\prime }$ and $H_{0}^{\prime }$ in order to get an integrable system as
well. It may happen that some of the constraints obey (\ref{25}) identically. In this case these constraints can be related to first-class
constraints in the Hamiltonian point of view. If all $H_{\bar{\beta}}^{\prime }$ happen to be in this condition there is nothing left to do at
the classical level. The system is completely integrable with the dynamics of GB and the the dynamical equations must be taken from (\ref{24}). Notice
that this differential defines an $n-m+1$ parameter evolution, in which $H_{\bar{\beta}}^{\prime }$ are the generators and $x^{\bar{\beta}}$ remains
as parameters.

However, if the equations (\ref{25}) are not identically satisfied, they will result in relations between the variables themselves, which indicates
that the system was not completed by the previous analysis. In this case these relations must be taken as new constraints that must be added to the
former $H_{\bar{\beta}}^{\prime }$ and the analysis must be remade until no relations come out of (\ref{25}).

\section{The First-Order EH Action}

The first-order form of the EH action in $n$ even dimensions is given by the functional%
\begin{equation}
S_{n}=\int d^{\left( d\right) }x\ \sqrt{-g}g^{\alpha \beta }\left[ \Gamma
_{\alpha \beta ,\lambda }^{\lambda }-\Gamma _{\alpha \lambda ,\beta
}^{\lambda }+\Gamma _{\sigma \lambda }^{\lambda }\Gamma _{\alpha \beta
}^{\sigma }-\Gamma _{\sigma \beta }^{\lambda }\Gamma _{\alpha \lambda
}^{\sigma }\right] ,  \label{26}
\end{equation}%
also called the Palatini action. The $g^{\alpha \beta }$ variables are components of the metric, and $\Gamma _{\alpha \beta }^{\gamma }$ are the
components of the affine connection, which we will consider symmetric in the lower indexes. The variational principle with fixed boundary of this
action is often called metric-affine variation, \emph{ie} metric and affine connection are varied independently, in contrast with the pure metric
variation, which is taken by the variation of the metric alone.

The compatibility of both approaches has been analyzed for some time for the General Relativity (GR)\cite{11}. Although the connection is chosen to be
symmetric in both cases, in the metric variation we also choose the metricity condition, $Dg=0$, so that the connection is related to the metric
by the Christoffel's symbols. In the metric-affine variation, metric and connection are assumed to be independent, and if we need compatibility with
GR we should impose the metricity condition as a constraint in the Lagrangian, which is a rather difficult task, since it is a constraint that
has derivatives of the $g$ fields. In this work we will avoid this problem by not forcing the compatibility between the metric and metric-affine
theories, working with the action (\ref{26}) without the metricity as a constraint. By doing so we notice that GR is a particular case of the theory
since some of the field equations of the metric-affine approach does not fix the connection as the Christoffel's symbols.

The theory in two dimensions was believed to have no canonical description \cite{jackiw}, since Einstein's equations become identically satisfied,
and the EH action would be a pure surface term. Although a surface term in the action gives rise to trivial equations of motion, the converse is not,
in general, true. In fact, the EH action in two dimensions is not a surface term, but has a part that depends on non-diagonal components of the metric
\cite{grav2}. It is possible to choose a coordinate system in which the metric is diagonal, because of the fact that any $\left( 1+1\right) $
dimensional surface is conformally flat \cite{fomenko}. However, if we desire to obey general covariance, we are actually able to define
appropriate canonical momenta for the metric fields.

We are able to write the Lagrangian density as%
\begin{equation}
\mathcal{L}=-\ \Omega _{\gamma }^{\alpha \beta }\dot{\Gamma}_{\alpha \beta
}^{\gamma }-\Psi _{\alpha \beta }\dot{g}^{\alpha \beta }-\mathcal{H}_{0},
\label{27}
\end{equation}%
which has weight 1. Let us define the symmetric symbol $\Delta _{\mu \nu
}^{\alpha \beta }\equiv 1/2\left( \delta _{\mu }^{\alpha }\delta _{\nu
}^{\beta }+\delta _{\mu }^{\beta }\delta _{\nu }^{\alpha }\right) $. Then,
the functions that appear in the Lagrangian can be written by
\label{F1}
\begin{eqnarray}
\Psi _{\alpha \beta } &=&0,  \label{psi} \\
\Omega _{\gamma }^{\alpha \beta } &=&\sqrt{-g}\left[ -g^{\alpha \beta
}\delta _{\gamma }^{0}+\Delta _{\pi \gamma }^{\alpha \beta }g^{0\pi }\right]
,  \label{omega} \\
\mathcal{H}_{0} &=&\sqrt{-g}g^{\alpha \beta }\left[ \Delta _{\alpha \beta
}^{i\rho }\Gamma _{\rho \lambda ,i}^{\lambda }-\Gamma _{\alpha \beta
,i}^{i}-\Gamma _{\sigma \lambda }^{\lambda }\Gamma _{\alpha \beta }^{\sigma
}+\Gamma _{\sigma \alpha }^{\lambda }\Gamma _{\beta \lambda }^{\sigma }%
\right] .  \label{H}
\end{eqnarray}%
In this expression $\mathcal{H}_{0}$ stands for the Hamiltonian density of the system. Since the Lagrangian depends only of the first derivatives of
the fields, (\ref{26}) is a first-order action on the variables $\Gamma_{\alpha \beta }^{\gamma }$.

Another way to analyze the theory behind the action (\ref{26}) is using the metric density $h^{\alpha \beta }\equiv \sqrt{-g}g^{\alpha \beta }$ as
variables in place of the metric. In $d>2$ dimensions both ways are completely equivalent, but it is not true for $d=2$. If we take the
determinant of the density we have
\begin{equation}
h\equiv \det h^{\alpha \beta }=-(-g)^{\frac{d-2}{2}}.  \label{31}
\end{equation}%
When $d=2$ we have $h=-1$, and it is not possible to write the determinant of the metric $g$ in function of $h$. The equation $h=-1$ is actually a
constraint between the $h$ variables, so we expect the gravitational field to have only two degrees of freedom in two dimensions when one considers the $h^{\alpha \beta }$ fields, instead of the three degrees of freedom if we use the $g^{\alpha \beta }$ variables. In this case it is impossible to write the metric in terms of the density, and a pure metric density approach is impossible to be performed, since we cannot write the Christoffel symbols and it derivatives in terms of $h^{\alpha \beta }$. Therefore, the only possible analysis in two dimensions that involves the metric density is the Palatini's approach.

The set of constraints of the theory is given by
\label{con1}
\begin{eqnarray}
\phi _{0} &\equiv &p_{0}+\mathcal{H}_{0}=0,  \label{p0} \\
\phi _{\alpha \beta } &\equiv &\pi _{\alpha \beta }=0,  \label{p1} \\
\phi _{\gamma }^{\alpha \beta } &\equiv &\pi _{\gamma }^{\alpha \beta
}+\Omega _{\gamma }^{\alpha \beta }=0,  \label{p2}
\end{eqnarray}%
and for this set we need to test the integrability conditions.

\subsection{Analysis of the $M$ matrix}

As we saw in section 2.2 the analysis of the integrability condition can be made on the analysis of the zero modes of the $M$ matrix defined in
(\ref{m1}), then we should define our fundamental PB relations as
\begin{eqnarray*}
\{g^{\alpha \beta }(x),\pi _{\mu \nu }(y)\}=\Delta _{\mu \nu }^{\alpha \beta
}\delta ^{\left( d-1\right) }\left( x-y\right), \\
\{\Gamma_{\alpha \beta }^{\lambda }(x),\pi _{\gamma }^{\mu \nu }(y)\}=\delta
_{\gamma }^{\lambda }\Delta _{\alpha \beta }^{\mu \nu }\delta ^{\left(
d-1\right) }\left( x-y\right) .
\end{eqnarray*}%
The $M$ matrix in which we are interested is the matrix of the PB between the constraints (\ref{p1}) and (\ref{p2}). Using the result%
\begin{equation}
\delta \sqrt{-g}=-\frac{1}{2}\sqrt{-g}g_{\varepsilon \eta }\delta
g^{\varepsilon \eta },  \label{dg}
\end{equation}%
the brackets which we are interested have the form%
\begin{eqnarray*}
\left\{ \phi _{\alpha \beta },\phi _{\mu \nu }\right\} &=&0,\ \ \ \ \ \ \
\left\{ \phi _{\gamma }^{\alpha \beta },\phi _{\lambda }^{\mu \nu }\right\}
=0 \\
\left\{ \phi _{\alpha \beta },\phi _{\lambda }^{\mu \nu }\right\} &=&\frac{1%
}{2}\sqrt{-g}\left[ g_{\alpha \beta }g^{\rho \tau }-2\Delta _{\alpha \beta
}^{\rho \tau }\right] \left[ -\Delta _{\rho \tau }^{\mu \nu }\delta
_{\lambda }^{0}+\Delta _{\rho \tau }^{0\sigma }\Delta _{\sigma \lambda
}^{\mu \nu }\right] \delta ^{\left( d-1\right) }\left( x-y\right) .
\end{eqnarray*}

Therefore, we have the matrix%
\begin{equation}
M\left( x,y\right) \equiv \left(
\begin{array}{cc}
0 & A\left( x,y\right) \\
-A^{T}\left( x,y\right) & 0%
\end{array}%
\right) ,  \label{M3}
\end{equation}%
in which we set $A\equiv \left\{ \phi _{\alpha \beta },\phi _{\lambda }^{\mu\nu }\right\} $. If we take the system in two dimensions we will deal with
the nine variables $\left( g^{00},g^{01},g^{11},\Gamma _{00}^{0},\Gamma_{00}^{1},\Gamma _{01}^{0},\Gamma _{01}^{1},\Gamma _{11}^{0},\Gamma
_{11}^{1}\right) $. Then, the matrix $A$ can be written by%
\begin{equation}
A=\frac{1}{4}\sqrt{-g}\left(
\begin{array}{cccccc}
0 & 0 & -g_{00}g^{01} & g_{00}g^{00}-2 & -2g_{00}g^{11} & 2g_{00}g^{01} \\
0 & 0 & g_{00}g^{00} & g_{01}g^{00} & -2g_{01}g^{11} & -2g_{00}g^{00} \\
0 & 0 & -g_{11}g^{01} & g_{11}g^{00} & 4-2g_{11}g^{11} & 2g_{11}g^{01}%
\end{array}%
\right) \delta \left( x-y\right) .  \label{A3}
\end{equation}

The matrix $M$ is singular. It is clear that two zero modes are related to the fact that the constraints $\phi _{0}^{00}$ and $\phi _{1}^{00}$ are in
involution with the set of HJPDE of the system. There must be another modes, since a skew-symmetric matrix of odd dimension has to be singular as
well. We can see that the constraints $\phi _{0}^{01}$ and $\phi _{1}^{11}$ are not linearly independent on the matrix space. To apply the
integrability conditions we are looking for the basis of the regular modes of the $M$ matrix, that corresponds to the largest linearly independent set
of non-involutive constraints, in order to build the GB of the system. Let us exclude $\phi _{1}^{11}$ of this set and build the matrix of the PB of
constraints $\left( \phi _{00},\phi _{01},\phi _{11},\phi _{0}^{01},\phi_{1}^{01},\phi _{0}^{11}\right) $. It can be written by%
\begin{equation}
M^{\prime }\left( x,y\right) \equiv \frac{1}{4}\sqrt{-g}\left(
\begin{array}{cc}
0_{3\times 3} & B_{3\times 3}\left( x\right) \\
-\left( B^{T}\right) _{3\times 3}\left( x\right) & 0_{3\times 3}%
\end{array}%
\right) \delta \left( x-y\right) ,  \label{M'3}
\end{equation}%
where%
\begin{equation}
B \equiv \left(
\begin{array}{ccc}
-g_{00}g^{01} & g_{00}g^{00}-2 & -2g_{00}g^{11} \\
1-g_{01}g^{01} & g_{01}g^{00} & -2g_{01}g^{11} \\
-g_{11}g^{01} & g_{11}g^{00} & 4-2g_{11}g^{11}%
\end{array}%
\right) .  \label{B3}
\end{equation}

Because of the previous analysis we expect $M^{\prime }$ to be regular, and it is sufficient to calculate the inverse of the matrix $B\left( x\right)
$. However, the determinant of this matrix is given by%
\[\det B =4\left( 2-g_{\alpha \beta }g^{\alpha \beta }\right) .\]
It is identically zero in two dimensions, because of the fact that $g_{\alpha \beta }g^{\alpha \beta }=d.$ This problem is due to the fact that,
introducing the fields $g_{\alpha\beta}$ by (\ref{dg}), the expression $g_{\alpha \beta }g^{\alpha \beta }-2=0$ is actually a constraint. In this way, we expect that exists at least another involutive constraints in the theory. Our procedure should be to insert this constraint on the original
Lagrangian density with a Lagrange multiplier and remake the analysis considering the multiplier as another arbitrary field. Let us just try to choose
a more restrict sub-matrix from $M^{\prime }$ instead.

Let us consider the subset of constraints $H_{a}^{\prime }=\left( \phi_{01},\phi _{11},\phi _{0}^{01},\phi _{1}^{01}\right) $. The matrix $%
M_{ab}^{\prime \prime } \equiv \left\{ H_{a}^{\prime}\left( x\right) ,H_{b}^{\prime }\left( y\right) \right\} $ can be written by%
\begin{equation}
M^{\prime \prime }\left( x,y\right) \equiv \frac{1}{4}\sqrt{-g}g_{11}\left(
\begin{array}{cc}
0_{2\times 2} & B^{\prime }\left( x\right) \\
-\left( B^{\prime }\right) ^{T}\left( x\right) & 0_{2\times 2}%
\end{array}%
\right) \delta \left( x-y\right) ,  \label{M''}
\end{equation}%
in which%
\begin{equation}
B^{\prime }\left( x\right) \equiv \left(
\begin{array}{cc}
g^{11} & -g^{01} \\
-g^{01} & g^{00}%
\end{array}%
\right) .  \label{B'}
\end{equation}%
For this matrix we have $\det B^{\prime }\left( x\right) =g_{11}g^{00},$ and both $B^{\prime }\left( x\right) $ and $M^{\prime \prime }\left(
x,y\right) $ are regular matrices.

The inverse of $M^{\prime \prime }\left( x,y\right) $ is given by%
\begin{equation}
\left( M^{\prime \prime }\right) ^{-1} =\frac{4}{\sqrt{-g}}%
\frac{1}{g^{00}}\left(
\begin{array}{cc}
0_{2\times 2} & -\left( B^{\prime T}\right) ^{-1} \\
\left( B^{\prime }\right) ^{-1} & 0_{2\times 2}%
\end{array}%
\right) \delta \left( x-y\right) .  \label{invM''}
\end{equation}%
Since $\left( B^{\prime T}\right) ^{-1}=\left( B^{\prime -1}\right) ^{T},$
all we have to do is calculate the inverse of $B^{\prime }\left( x\right) ,$
which is given by%
\begin{equation}
\left( B^{\prime }\right) ^{-1} =\left(
\begin{array}{cc}
g^{00} & g^{01} \\
g^{01} & g^{11}%
\end{array}%
\right) .  \label{invB'}
\end{equation}

With this matrix, and following the development of the HJ formalism, we are able to build the GB%
\begin{equation}
\left\{ F,G\right\} ^{\ast }=\left\{ F,G\right\} -\left\{ F,H_{a}^{\prime
}\right\} \left( M^{\prime \prime -1}\right) _{ab}\left\{ H_{b}^{\prime
},G\right\} ,  \label{GB3}
\end{equation}%
where a double integration is implicit on the second term of the right hand side.

\subsection{Fundamental GB and algebra of generators}

These GB give rise to the following nonzero fundamental brackets:
\begin{eqnarray}
\left\{ g^{\alpha \beta },\Gamma _{\mu \nu }^{\lambda }\right\} ^{\ast } &=& \frac{4}{\sqrt{-g}}\frac{\Delta _{\mu \nu }^{01}}{g^{00}}\left[ -\delta_{0}^{\lambda }\Delta _{01}^{\alpha \beta }g^{00}-\left( \delta_{1}^{\lambda }\Delta _{01}^{\alpha \beta }+\delta _{0}^{\lambda }\Delta
_{11}^{\alpha \beta }\right) g^{01}-\delta _{1}^{\lambda }\Delta
_{11}^{\alpha \beta }g^{11}\right] \delta \left( x-y\right),  \label{a3} \\
\left\{ g^{\alpha \beta },\pi _{\mu \nu }\right\} ^{\ast } &=&\frac{%
g^{\alpha \beta }}{g^{00}}\Delta _{\mu \nu }^{00}\delta \left( x-y\right) ,
\label{b3} \\
\left\{ \Gamma _{\alpha \beta }^{\gamma },\pi _{\lambda }^{\mu \nu }\right\}
^{\ast } &=&\delta _{\lambda }^{\gamma }\Delta _{\alpha \beta }^{\mu \nu
}\delta \left( x-y\right) .  \label{c3}
\end{eqnarray}%
The GB between the constraints are given by
\[\left\{ \phi _{\alpha \beta },\phi _{\mu \nu }\right\} ^{\ast }=0,\ \ \ \ \
\left\{ \phi _{\gamma }^{\alpha \beta },\phi _{\lambda }^{\mu \nu }\right\}
^{\ast }=0,\ \ \ \ \ \left\{ \phi _{\alpha \beta },\phi _{\lambda }^{\mu \nu
}\right\} ^{\ast }=0,\]
so the set of constraints $\left( \phi _{\alpha \beta },\phi _{\lambda}^{\mu \nu }\right) $ is in involution.

The integrability conditions of the constraints $\left( \phi _{00},\phi_{0}^{11},\phi _{0}^{00},\phi _{1}^{00},\phi _{1}^{11}\right) $ are yet to
be tested, by equations (\ref{25}). The brackets
\begin{eqnarray*}
\left\{ \phi _{\alpha \beta },\phi _{0}\right\} ^{\ast }&=&-\Delta _{\alpha
\beta }^{00}\sqrt{-g}\frac{g^{\mu \nu }}{g^{00}}\left[ 1-\frac{1}{2}d\right] %
\left[ \Delta _{\mu \nu }^{1\rho }\Gamma _{\rho \lambda ,1}^{\lambda
}-\Gamma _{\mu \nu ,1}^{1}-\Gamma _{\sigma \lambda }^{\lambda }\Gamma _{\mu
\nu }^{\sigma }+\Delta _{\mu \nu }^{\rho \tau }\Gamma _{\sigma \rho
}^{\lambda }\Gamma _{\tau \lambda }^{\sigma }\right] \delta \left( x-y\right)
\end{eqnarray*}%
are actually zero, and $\phi _{00}$ has its integrability identically satisfied.

We also have the equations%
\[\left\{ \phi _{\gamma }^{\alpha \beta },\phi _{0}\right\} ^{\ast }=0.\]
These ones are not identically satisfied. They result in the following additional conditions on the metric fields:
\begin{eqnarray}
\chi _{1} &\equiv &\partial _{1}\left( \sqrt{-g}g^{01}\right) +\sqrt{-g}\left[g^{00}\Gamma _{01}^{1}+g^{11}\Gamma _{11}^{0}\right]=0,  \label{x1} \\
\chi _{2} &\equiv &-\partial _{1}\left( \sqrt{-g}g^{00}\right) -\sqrt{-g}\left[g^{00}\left( \Gamma _{01}^{0}-\Gamma _{11}^{1}\right) -2g^{01}\Gamma _{11}^{0}\right]=0,  \label{x2} \\
\chi _{3} &\equiv &\partial _{1}\left( \sqrt{-g}g^{11}\right) -\sqrt{-g}\left[g^{11}\left( \Gamma _{01}^{0}-\Gamma _{11}^{1}\right) +2g^{01}\Gamma _{01}^{1}\right]=0,  \label{x3} \\
\chi _{4} &\equiv &\frac{1}{g^{00}}\left[ g^{11}\chi _{1}-2g^{01}\chi _{3}%
\right] =0.  \label{x4}
\end{eqnarray}%
They should be considered as new constraints of the theory.

The condition $\chi _{4}$ is not linearly independent of the others, so we should analyze the integrability of the set $\left( \phi
_{00},\phi_{0}^{11},\phi _{0}^{00},\phi _{1}^{00},\phi_{1}^{11},\chi _{1},\chi _{2},\chi _{3}\right) $, which involves the calculation of the GB of
the constraints. Let us set $\chi_{A}\equiv \left( \chi _{1},\chi _{2},\chi _{3}\right) $, we have the results
\begin{equation}
\left\{ \chi _{A},\phi _{\alpha \beta }\right\} ^{\ast }=0,\ \ \ \ \ \left\{
\chi _{A},\phi _{\lambda }^{\mu \nu }\right\} ^{\ast }=0 ,  \label{gbx1}
\end{equation}%
and%
\begin{eqnarray*}
\left\{ \chi _{1},\chi _{2}\right\} ^{\ast }=-\chi _{2}\delta \left(
x-y\right) ,\\
\left\{ \chi _{1},\chi _{3}\right\} ^{\ast }=\chi
_{3}\delta \left( x-y\right) , \\
\left\{ \chi _{2},\chi _{3}\right\}
^{\ast }=2\chi _{1}\delta \left( x-y\right) .
\end{eqnarray*}

Of course, there still remain the conditions $\left\{ \chi _{A},\phi_{0}\right\} ^{\ast }=0$ to be tested. However, it is easy to show that
the canonical Hamiltonian of the system is a linear combination of these constraints:
\begin{equation}
\mathcal{H}_{0}=\left( \Gamma _{01}^{1}-\Gamma _{00}^{0}\right) \chi
_{1}-\Gamma _{00}^{1}\chi _{2}-\Gamma _{01}^{0}\chi _{3},  \label{h03}
\end{equation}%
then, these conditions are also identically satisfied. Therefore, the complete set of constraints, $ \phi _{0,}\phi _{\alpha \beta },\phi_{\gamma
}^{\alpha \beta },\chi _{A} $ are in involution with the GB operation, and hence, full integrability is achieved.

We are allowed to perform the transformation
\begin{equation}
\sigma _{1}\equiv \frac{1}{2}\left( \chi _{3}+\chi _{2}\right) ,\ \ \ \ \
\sigma _{2}\equiv \frac{1}{2}\left( \chi _{3}-\chi _{2}\right) ,\ \ \ \ \
\sigma _{3}\equiv \chi _{1},  \label{sigma}
\end{equation}%
which gives the GB fundamental relations
\begin{equation}
\left\{ \sigma _{1},\sigma _{2}\right\} ^{\ast }=\sigma _{3},\ \ \ \ \left\{
\sigma _{3},\sigma _{1}\right\} ^{\ast }=\sigma _{2},\ \ \ \ \left\{ \sigma
_{2},\sigma _{3}\right\} ^{\ast }=-\sigma _{1}.  \label{so21}
\end{equation}
With respect to the GB operation, this is the algebra of the $SO\left(2,1\right) $ group \cite{grav1}.

Considering the metric density $h^{\alpha \beta }\equiv \sqrt{-g}g^{\alpha\beta }$, and the derivative
\begin{equation}
D_{\gamma }h^{\alpha \beta }\equiv \partial _{\gamma }h^{\alpha \beta
}+2\Delta _{\mu \nu }^{\alpha \beta }h^{\mu \sigma }\Gamma _{\gamma \sigma
}^{\nu }-h^{\alpha \beta }\Gamma _{\gamma \mu }^{\mu },  \label{D}
\end{equation}%
which is the covariant derivative for a density tensor of rank 2, the constraints $\chi _{A}$ can be written by
\begin{equation}
D_{1}h^{\alpha \beta }=0,  \label{D1}
\end{equation}%
which is a geometric relation on the densities.

\subsection{Field equations}

The HJ formalism, through the GB structure, naturally separates involutive and non-involutive constraints with respect to the PB operation, and in our
system we have the set $\left( \phi _{00},\phi _{0}^{11},\phi _{0}^{00},\phi_{1}^{00},\phi _{1}^{11},\chi _{A}\right) $ of involutive constraints, as
well as the set $\left( \phi _{01},\phi _{11},\phi _{0}^{01},\phi_{1}^{01}\right) $ of non-involutive ones. The GB of non-involutive constraints with
any other phase space function are zero by construction, so we are allowed to write
\begin{eqnarray}
\delta F&=&\left\{ F,\phi _{0}\right\} ^{\ast }\delta t+\left\{ F,\phi_{\alpha \beta }\right\} ^{\ast }\delta g^{\alpha \beta }+\left\{ F,\phi
_{\gamma }^{\alpha \beta }\right\} ^{\ast }\delta \Gamma _{\alpha \beta}^{\gamma }+\left\{ F,\chi _{A}\right\} ^{\ast }\delta \omega ^{A},
\label{dF}
\end{eqnarray}%
as the fundamental variation that defines the dynamics of the system. To include the conditions $\chi_A=0$ we had to introduce the parameters
$\omega^{A}$, which are arbitrary fields. Integration is assumed on the terms of the right hand side.

The field equations for the variables $g^{\mu \nu }$ are given by
\begin{eqnarray*}
\delta g^{\mu \nu }&=&\left\{ g^{\mu \nu },\phi _{0}\right\} ^{\ast }\delta
t+\left\{ g^{\mu \nu },\phi _{\alpha \beta }\right\} ^{\ast }\delta
g^{\alpha \beta }+\left\{ g^{\mu \nu },\phi _{\gamma }^{\alpha \beta
}\right\} ^{\ast }\delta \Gamma _{\alpha \beta }^{\gamma }+\left\{ g^{\mu
\nu },\chi _{A}\right\} ^{\ast }\delta \omega ^{A},
\end{eqnarray*}%
that result in the relations
\begin{eqnarray*}
\delta g^{00} &=&\delta g^{00}, \\
\partial _{0}g^{01} &=&-g^{00}\Gamma _{00}^{1}-g^{11}\Gamma
_{01}^{0}-g^{01}\left( \Gamma _{01}^{1}-\Gamma _{00}^{0}\right) +2\frac{%
g^{01}g^{01}}{g^{00}}\Gamma _{01}^{0} \\
&&+\frac{g^{01}}{g^{00}}\partial
_{0}g^{00}-g^{01}\dot{\omega}^{1}+g^{00}\dot{\omega}^{2}+\frac{1}{g^{00}}\left[
g^{00}g^{11}-2g^{01}g^{01}\right] \dot{\omega}^{3}, \\
\partial _{0}g^{11} &=&-2g^{11}\left( \Gamma _{01}^{1}-\Gamma
_{00}^{0}\right) -2g^{01}\Gamma _{00}^{1}+2\frac{g^{01}g^{11}}{g^{00}}\Gamma
_{01}^{0}+\frac{g^{11}}{g^{00}}\partial _{0}g^{00} \\
&&-2g^{11}\dot{\omega}^{1}+2g^{01}\dot{\omega}^{2}-2\frac{g^{01}g^{11}}{%
g^{00}}\dot{\omega}^{3},
\end{eqnarray*}%
where we used the dot to indicate partial differentiation with respect to $t$. The variation $\delta g^{00}$ is still arbitrary. It is possible to
show that, if we choose the parameters $\omega ^{A}$ as independent of $t$, the above field equations can be summarized by%
\begin{equation}
D_{0}h^{\alpha \beta }=0.  \label{D0}
\end{equation}

Together, equations (\ref{D1}) and (\ref{D0}) give us the following
geometric relation,
\begin{equation}
D_{\gamma }h^{\alpha \beta }=0,  \label{Dgamma}
\end{equation}
which is shown in \cite{Deser} by direct metric density-affine variation of the EH action. As direct consequence of these equations we have that
the space-time in two dimensions does not obey metricity, and the affine connection cannot be fixed to be the Christoffel's symbols, due to the fact
that the trace of the connection is not well defined.

Equations for $\Gamma _{\mu \nu }^{\lambda }$ are given by
\[\delta \Gamma _{\mu \nu }^{\lambda }=\left\{ \Gamma _{\mu \nu }^{\lambda
},\phi _{0}\right\} ^{\ast }\delta t+\left\{ \Gamma _{\mu \nu }^{\lambda
},\phi _{\gamma }^{\alpha \beta }\right\} ^{\ast }\delta \Gamma _{\alpha
\beta }^{\gamma }+\left\{ \Gamma _{\mu \nu }^{\lambda },\chi _{A}\right\}
^{\ast }\delta \omega ^{A},\]
which gives%
\begin{eqnarray*}
\Gamma _{01,0}^{0} &=&\Gamma _{00,1}^{0}-\Gamma _{01,1}^{1}+\Gamma
_{11,0}^{1}+2\Gamma _{00}^{1}\Gamma _{11}^{0}-2\Gamma _{01}^{0}\Gamma
_{01}^{1} \\
&&+2\frac{g^{01}}{g^{00}}\left[ \Gamma _{01,1}^{0}-\Gamma _{11,0}^{0}+\Gamma
_{01}^{0}\left( \Gamma _{01}^{0}-\Gamma _{11}^{1}\right) +\Gamma
_{11}^{0}\left( \Gamma _{01}^{1}-\Gamma _{00}^{0}\right) \right] , \\
\Gamma _{01,0}^{1} &=&\Gamma _{00,1}^{1}-\Gamma _{01}^{1}\left( \Gamma
_{01}^{1}-\Gamma _{00}^{0}\right) -\Gamma _{00}^{1}\left( \Gamma
_{01}^{0}-\Gamma _{11}^{1}\right) \\
&&+\frac{g^{11}}{g^{00}}\left[ -\Gamma _{11,0}^{0}+\Gamma _{01,1}^{0}+\Gamma
_{01}^{0}\left( \Gamma _{01}^{0}-\Gamma _{11}^{1}\right) +\Gamma
_{11}^{0}\left( \Gamma _{01}^{1}-\Gamma _{00}^{0}\right) \right] ,
\end{eqnarray*}%
where the variations $\delta \omega ^{A}$ are set to be zero. The remaining equations are just%
\[\delta \Gamma _{00}^{\lambda }=\delta \Gamma _{00}^{\lambda },\ \ \ \ \ \ \
\ \ \ \ \ \ \ \ \ \ \delta \Gamma _{11}^{\lambda }=\delta \Gamma
_{11}^{\lambda }.\]

The equations for $\Gamma _{01,0}^{0}$ and $\Gamma _{01,0}^{1}$ can be written with the components of the Ricci tensor
\begin{eqnarray}
g^{00}R_{\left( 01\right) }+g^{01}R_{11} =0,  \label{3aa} \\
g^{00}R_{00}-g^{11}R_{11} =0,  \label{4aa}
\end{eqnarray}%
in which $R_{\left( \alpha \beta \right) }\equiv 1/2\left( R_{\alpha \beta}+R_{\beta \alpha }\right) $. We must notice that, because the trace of the
connection is an arbitrary 1-form in two dimensions, the Ricci tensor is not symmetric. The tensor solution of this system has only one arbitrary
independent component, as expected. Let us take a look at the Einstein's equations in the absence of sources
\begin{equation}
R_{\mu \nu }-\frac{1}{2}g_{\mu \nu }R=0.  \label{ee}
\end{equation}%
If we make the contraction of this equation with $g^{\gamma \mu }$ it is easy to see that the resulting equations are just equivalent to (\ref{3aa},\ref{4aa}).
It means that the field equations obtained are equivalent to a partially contracted set of Einstein's equations, which in the case of absence of
sources gives no lack of information.

Finally, let us talk a little about the HJ formalism applied to the same problem, but now using as dynamical variables the metric density
$h^{\alpha\beta }$ and the affine connection $\Gamma _{\mu \nu }^{\gamma }.$ As we saw in equation (\ref{31}), the two dimensional action is
constrained by the requirement that the determinant of the metric density is $-1.$ Because of the similar operational procedure used here, the HJ and
Dirac approaches using the density differ only in minor details, and the same results of \cite{grav1} are obtained. In particular, the $M$ matrix is
not field-dependent, which in fact simplifies greatly the procedure. The set of non-involutive constraints is different, but because of the fact that
the constraint $\chi_{4}=0$ is a combination of other involutive constraints, the counting of the degrees of freedom of the metric and the metric
density is the same. The set of independent involutive constraints also obeys the algebra of (\ref{so21}). The only noticeable difference lies on the
field equations, that in this case are equivalent to the fully contracted Einstein's equations, \emph{ie} Einstein's equations (\ref{ee}) contracted
with the metric $g^{\mu \nu}.$

\section{Final remarks}

In this work we have analyzed the two dimensional Einstein-Hilbert action within the Hamilton-Jacobi theory for constrained systems. It is already
known in the literature that General Relativity in two dimensions has several features which are not present in the theory in higher dimensions, as the
fact that the metric-affine variation (the Palatini's action) is not equivalent to the purely metric description, since metricity is not obeyed, and
the affine connection depends on arbitrary vector fields \cite{Deser}. Einstein's equation are trivial, as the metric is always conformal to
Minkowski one, what implies that no real gravitational field exists. However, it is possible to define canonical conjugate momenta for the metric
fields, provided that non-diagonal terms of the metric are allowed. Therefore, a constraint analysis can be performed.

We have chosen to work with the metric $g^{\alpha \beta}$ in Palatini's approach, which did make the procedure more complicated, in order to avoid the
problem of considering the constraint $\det h^{\alpha \beta}=-1$ in the action. In Dirac formalism \cite{grav1} this is a first-class secondary
constraint that does not affect the dynamics of the system, but the number of degrees of freedom of the metric density is reduced to two independent
fields. This is clearly in concordance with the fact that the metric in two dimensions is always diagonalizable.

In the HJ approach we found a set of non-involutive constraints, which was eliminated from the theory with the introduction of Generalized brackets.
However, an odd situation occurred when we analyzed the regular modes of the $M$ matrix of the Poisson brackets of the non-involutive constraint:
because, in two dimensions, the identity $g_{\alpha\beta}g^{\alpha\beta}=2$ holds, there was a hidden involutive constraint in the formalism.
Choosing the matrix (\ref{M''}) we were able to build a unique GB.

This choice of GB has revealed the constraint $\phi_{00}$ as the hidden involutive constraint. This implies that the true degrees of freedom of the
metric components are actually the same as for the metric density. In this case we have changed from the problem of considering artificially the condition
$h+1=0$ in the action to the problem of an additional involutive constraint that had naturally arised. The supplementary integrability conditions
(\ref{25}), on the other hand, gives the four constraints (\ref{x1},\ref{x2},\ref{x3},\ref{x4}). One of them, $\chi_4$, resulted to be a linear combination of the others,
therefore being an irrelevant condition. Since the canonical Hamiltonian is also a linear combination of these constraints, we achieved the full
integrability of the system with the dynamics defined by the GB (\ref{GB3}).

Since the non-involutive constraints are eliminated by the GB, the remaining constraints are involutive, and they are actually Hamiltonian generators
of the evolution of the system. Among these generators there is a subset, $\chi_A$, that closes a local Lie algebra of the $SO(2,1)$ group. The
conditions $\chi_A=0$ are actually geometric relations: the spatial covariant conservation of the metric density of the system.

Calculating the characteristics equations of the theory we have found other relations on the metric, which, after a proper choice of the arbitrary
fields $\omega^A$, resulted to be the other part, the ``time'' component of the covariant conservation of the metric density. Therefore, we have
obeyed the condition $D_\gamma h^{\alpha\beta}=0$, which states that the space-time does not, generally, obeys metricity. The remaining field
equations are the ones that contain the dynamic of the theory, reproducing partially contracted Einstein's equations.

\section*{Acknowledgments}

MCB was supported by CAPES. BMP was partially supported by CNPq. PJP thanks the staff of IFI for the incentive and support.

\end{document}